\documentstyle[12pt]{article}

\def\1ad{\mbox{\normalsize $^1$}}
\def\2ad{\mbox{\normalsize $^2$}}
\def\3ad{\mbox{\normalsize $^3$}}
\def\4ad{\mbox{\normalsize $^4$}}
\def\5ad{\mbox{\normalsize $^5$}}
\def\6ad{\mbox{\normalsize $^6$}}
\def\7ad{\mbox{\normalsize $^7$}}
\def\8ad{\mbox{\normalsize $^8$}}
\def\makefront{\vspace*{1cm}\begin{center}
\def\newtitleline{\\ \vskip 5pt}
{\Large\bf\titleline}\\
\vskip 1truecm
{\large\bf\authors}\\
\vskip 5truemm
\addresses
\end{center}
\vskip 1truecm
{\bf Abstract:}
\abstracttext
\vskip 1truecm}
\begin {document}

\let\useblackboard=\iftrue
%

\def\beq{\begin{equation}}
\def\eeq{\end{equation}}
\def\beqa{\begin{eqnarray}}
\def\eeqa{\end{eqnarray}}

\def\xversim#1#2{\lower2.pt\vbox{\baselineskip0pt \lineskip-.5pt
x  \ialign{$\mxth#1\hfil##\hfil$\crcr#2\crcr\sim\crcr}}}

\def\beq{\begin{equation}}
\def\eeq{\end{equation}}
\def\beqa{\begin{eqnarray}}
\def\eeqa{\end{eqnarray}}

\newcommand{\ba}{\begin{array}}
\newcommand{\ea}{\end{array}}
\newcommand{\eq}{\begin{equation}}

\useblackboard
\typeout{If you do not have msbm (blackboard bold) fonts,}
\typeout{change the option at the top of the tex file.}
\font\blackboard=msbm10 scaled \magstep1
\font\blackboards=msbm7
\font\blackboardss=msbm5
\newfam\black
\textfont\black=\blackboard
\scriptfont\black=\blackboards
\scriptscriptfont\black=\blackboardss

\else

\fi

\newcommand{\eqn}{\begin{eqnarray}}
\newcommand{\enq}{\end{eqnarray}}
\newcommand{\eqa}{\begin{array}}
\newcommand{\ena}{\end{array}}
\newcommand{\en}{\end{equation}}

\newcommand{\no}{\nonumber}

\def\del{\partial}
\def\half{{1\over 2}}
\def\hhalf{{1\over 4}}
\def\Tr{{\rm Tr\ }}

\def\titleline{
Three-dimensional solutions of M-theory 
\newtitleline
on $S^1/Z_2 \times T_7$
}
\def\authors{
Gabriel Lopes Cardoso 
}
\def\addresses{
Institute for Theoretical Physics, Utrecht University\\
NL-3508 TA Utrecht, The Netherlands 
}
\def\abstracttext{
We review various three-dimensional solutions
in the low-energy description of M-theory on 
$S^1/Z_2 \times T_7$.  These solutions
have an eleven dimensional interpretation in terms
of intersecting M-branes.
}
\makefront
\section{Introduction}

Eleven dimensional supergravity, the low-energy effective theory of M-theory,
has four $1/2$-supersymmetric solutions:
the M-2-brane, the M-5-brane, the M-wave and
(in the $S^1$ Kaluza-Klein vacuum) the M-monopole.  
We will refer to them collectively as M-branes.
Intersections
of these four basic configurations give rise to solutions preserving
a smaller amount of supersymmetry
(for a review see for instance \cite{DuRaLu,Gaunt}
and references therein).  In the following, we will 
review various three-dimensional solutions in the low-energy 
description
of M-theory on $S^1/Z_2 \times T_7$.  These solutions 
have an eleven dimensional interpretation in terms
of orthogonally intersecting M-branes.

The eleven-dimensional space-time line element describing
a supersymmetric configuration of 
orthogonally intersecting M-branes is given in terms of harmonic
functions which depend on some of the overall transverse directions
\cite{Gaunt,ts,berg1}.
If these overall transverse directions do not include the eleven-th 
dimension $x_{11}$,
then its compactification on $S^1/Z_2 \times T_7$
down to three dimensions is also a solution
of the low-energy effective field theory
of ten-dimensional
heterotic string theory compactified on a seven-torus.
Such a 
solution is determined in terms of harmonic functions 
$H(z, \bar z) = f(z) + {\bar f}(\bar z)$, where
$z$ and $\bar z$ denote the spatial dimensions.
The static solutions presented in 
\cite{sen1,BBC,BC} are of this type.
If, on the other hand, the overall transverse directions include
the eleven-th dimension $x_{11}$, then compactifying the eleven-dimensional
configuration on $S^1/Z_2 \times T_7$ gives rise to a 
solution which 
is determined in terms of harmonic functions $H(x_{11}, z, \bar z)$.
An example of such a supergravity solution has recently been given 
\cite{BlauLoHull}
in the context
of M-theory compactified on $T_8$.  There, it was also shown
that this solution corresponds to a BPS state
with mass that goes like $1/g^3$.  Evidence for the 
existence of BPS states with masses that go like  $1/g^3$ or higher
inverse powers of the string coupling constant $g$  
in M-theory compactifications down to three (and lower)
dimensions has been emerging in studies of U-duality multiplets
of M-theory on tori \cite{ElGiKuRa,BlauLoHull}.

If the harmonic functions $H$ are taken to be independent of the
eleven-th  dimension, then the resulting three-dimensional 
solutions can be turned into finite energy solutions \cite{sen1,BC}
by utilizing a mechanism first discussed in the context of four-dimensional
stringy cosmic string solutions \cite{GSVY}.  For instance, the 
solutions presented in \cite{BC} have, at spatial infinity
($z \rightarrow \infty$), an
asymptotic behaviour corresponding to $f(z) \propto \ln z$.  At finite
distance these asymptotic solutions  become
ill-defined
and so  need to be modified.  The associated corrections are all 
encoded in $f(z)$.  They can be determined by requiring the 
solutions to have finite energy.  This requirement, together with the 
appropriate asymptotic behaviour, 
determines $f(z)$ to be given by 
$f(z) \propto j^{-1}(z)$.  
In section 2 we will review some of the three-dimensional
solutions constructed in \cite{BC}, namely those which have a 
ten-dimensional
interpretation in terms of 
a fundamental string, a wave and up to
three orthogonally intersecting NS $5$-branes as well as up to 
three Kaluza--Klein monopoles.
These solutions are labelled by an integer $n$ with $n=1,2,3,4$.
The energy $E$ carried by these (one-center)
solutions is given by $E = 2n \,{\pi \over 6}$
(in units where $8 \pi G_N =1$).

Another class of solutions constructed in \cite{BC} consists of solutions
carrying energies $E = n \frac{\pi}{6}$, where $n=1,2,3,4$. An example
of a solution with $E = \frac{\pi}{6}$ is given by a wrapped M-monopole.
In section 3 we briefly review this solution.  We then compare it
to the solution recently
discussed in \cite{BlauLoHull}, which can be constructed by considering a
periodic array
of M-monopoles along a transverse direction and by
identifying this transverse direction with the eleven-th
dimension.  This latter solution is thus specified by a harmonic
function $H(x_{11}, z, \bar z)$ 
on $S^1/Z_2 \times {\rm R^2}$.

\section{A class of finite energy solutions}

The effective low-energy field theory of the ten-dimensional heterotic string
compactified on a seven-dimensional torus is obtained from reducing the
ten-dimensional $N=1$ supergravity theory coupled to $U(1)^{16}$ super 
Yang--Mills multiplets 
 (at a generic point in the moduli space). 
The massless ten-dimensional bosonic fields are the metric $G^{(10)}_{MN}$, 
the antisymmetric tensor field $B^{(10)}_{MN}$, the $U(1)$ gauge 
fields $A^{(10)I}_M$
and the scalar dilaton $\Phi^{(10)}$ with $(0\leq M,N\leq 9,\quad 
 1\leq I\leq 16)$.
The field strengths are $F^{(10)I}_{MN}=\del_MA^{(10)I}_N-\del_NA^{(10)I}_M$ 
and $H^{(10)}_{MNP}=(\del_MB^{(10)}_{NP}-\half A^{(10)I}_MF^{(10)I}_{NP})+$
cyclic permutations of $M,N,P$.

The bosonic part of the ten-dimensional action is
\eqn
{\cal S} &\propto & \int d^{10}x\sqrt{-G^{(10)}}
e^{-\Phi^{(10)}}[{\cal R}^{(10)}
+G^{(10)MN}\del_M\Phi^{(10)}\del_N\Phi^{(10)}\no\\
&& \qquad\qquad\qquad -{1\over 12}H^{(10)}_{MNP}H^{(10)MNP}
-\hhalf F^{(10)I}_{MN}F^{(10)IMN}].
\enq
The reduction to three dimensions (see \cite{sen1} and references therein)
introduces 
the graviton
$g_{\mu\nu}$, the dilaton $\phi \equiv \Phi^{(10)} -\ln \sqrt{\det G_{mn}}\,$,
 with $G_{mn}$
the internal 7D metric, 30 $U(1)$ gauge fields 
$A^{(a)}_\mu\equiv (\,A_\mu^{(1)m},\, A^{(2)}_{\mu m},\, A_\mu^{(3)I}\,)
\quad (a=1,\dots,30,\;m=1,\dots ,7,\; I=1,\dots ,16)\;,$
where $\;A_\mu^{(1)m}\;$ are the 7 Kaluza--Klein gauge fields coming from 
the reduction of $G_{MN}^{(10)},\;A_{\mu m}^{(2)}\equiv B_{\mu m} 
+ B_{mn}A_\mu^{(1)n}+\half a_m^IA_\mu^{(3)I}\;$ are the 7 gauge fields coming
from the reduction of $\;B_{MN}^{(10)}\;$ and $\;A_\mu^{(3)I}\equiv A_\mu^I\
-a_m^IA_\mu^{(1)m}\;$ are the 16 gauge fields from  $A_M^{(10)I}$.
The field strengths $F_{\mu\nu}^{(a)}$ are given by 
$F_{\mu\nu}^{(a)}=\del_\mu A_\nu^{(a)}-\del_\nu A_\mu^{(a)}$. 
Finally, $\;B_{MN}^{(10)}\;$ induces the two-form field 
$\;B_{\mu\nu}\;$ with field strength $\;H_{\mu\nu\rho}=\del_\mu 
B_{\nu\rho}-\half A_\mu^{(a)}L_{ab}F_{\nu\rho}^{(b)}$+ cyclic permutations.
 
The bosonic part of the
three-dimensional action in the Einstein frame is then 
\eqn
{\cal S}&=&\hhalf\int d^3x\sqrt{-g}\bigl\{{\cal R}-g^{\mu\nu}
\del_\mu\phi\del_\nu\phi - 
{1\over 12} e^{-4\phi}g^{\mu\mu'}g^{\nu\nu'}g^{\rho\rho'}H_{\mu\nu\rho}
H_{\mu'\nu'\rho'}\no\\
&& - \hhalf e^{-2\phi}g^{\mu\mu'}g^{\nu\nu'}F_{\mu\nu}^{(a)}
(LML)_{ab}F_{\mu'\nu'}^{(b)}+ {1\over 8}g^{\mu\nu}\Tr (\del_\mu ML\del_\nu
ML)\big\}\;\;,
\label{act3d}
\enq
where $a=1,\dots,30.$  Here $M$ denotes a matrix comprising the scalar
fields $G_{mn},\,a_m^I$ and $B_{mn}$ \cite{mah,sen1}.

We now construct static solutions by setting the associated three-dimensional
Killing spinor equations to zero.  In doing so, we restrict ourselves to 
backgrounds with $H_{\mu\nu\rho}= 0$ and $a_m^I=0$.  It can be checked that
the resulting solutions to the Killing spinor equations
satisfy the equations of motion derived from (\ref{act3d}).
The associated three-dimensional Killing spinor equations in the Einstein frame
are \cite{BBC}
\eqn
\delta \chi^I&=&\half e^{-2\phi} F^{(3)I}_{\mu\nu}
\gamma^{\mu\nu}\varepsilon ,\no\\
\delta\lambda &=& -\half e^{-\phi}\del_\mu \{\phi+\ln\det e_m^a\} 
\gamma^\mu\otimes{\bf I}_8\,\varepsilon 
+\hhalf e^{-2\phi}[-B_{mn}F^{(1)n}_{\mu\nu}+F^{(2)}_{\mu\nu m} 
]\gamma^{\mu\nu}\gamma^4\otimes\Sigma^m\varepsilon\no\\
&&+\hhalf e^{-\phi}\del_\mu B_{mn}\gamma^\mu\otimes\Sigma^{mn}
\varepsilon\;\;,\no\\
\delta{\psi}_\mu &=&\del_\mu\varepsilon + {1\over 4}\omega_{\mu\alpha\beta}
\gamma^{\alpha\beta}\varepsilon 
+\hhalf (e_{\mu\alpha}e_\beta^\nu \!-\!e_{\mu\beta}e_\alpha^\nu)
\del_\nu\phi\gamma^{\alpha\beta}\varepsilon \no\\
&& + {1\over 8}(e^n_a\del_\mu 
e_{nb}\!-\!e^n_b\del_\mu e_{na}){\bf I}_4\otimes\Sigma^{ab}\varepsilon 
\!-\!{1\over 4}e^{-\phi}[e^m_a F^{(2)}_{\mu\nu(m)} 
- e_{ma} F^{(1)m}_{\mu\nu}]\gamma^\nu\gamma^4\otimes\Sigma^a\varepsilon \no\\
&& \!-\!{1\over 8}\del_\mu B_{mn}\! \;{\bf I}_4
\otimes\Sigma^{mn}\varepsilon 
+\hhalf e^{-\phi}B_{mn}F_{\mu\nu}^{(1)n}\;\gamma^\nu\gamma^4\otimes
\Sigma^m\varepsilon\;\;,\no\\
\delta\psi_d &=&-{1\over 4}e^{-\phi}(e_d^m\del_\mu e_{ma}\!+\!e_a^m\del_\mu 
e_{md})\gamma^\mu\gamma^4\otimes\Sigma^a\varepsilon
+{1\over 8}e^{-2\phi}e^m_d B_{mn}F^{(1)n}_{\mu\nu} \;
\gamma^{\mu\nu}\varepsilon\no\\
&&+{1\over 4}e^{-\phi}e^m_de^n_a \del_\mu B_{mn}\gamma^\mu
\gamma^4\otimes\Sigma^a\varepsilon
-{1\over 8}e^{-2\phi}[\;e_{md}F_{\mu\nu}^{(1)m}
+e_d^mF^{(2)}_{\mu\nu m}\;]\gamma^{\mu\nu}\varepsilon\;\;,\label{killing}
\enq
where $\delta\psi_d\equiv e_d^m\delta\psi_m$ denotes the variation of the
internal gravitini.  Here we have performed a $3+7$ split of the 
ten-dimensional gamma matrices \cite{BBC}.

In the following, we review a particular class of static 
solutions to the Killing spinor equations (\ref{killing}) constructed in 
\cite{BBC,BC}.  The solutions in this class are labelled by an integer $n$
($n=1,2,3,4$).  The associated space-time line element is given by
\beqa
ds^2 = 
-dt^2 + H^{2n} d\omega d{\bar \omega} \;\;\;, \;\;\; H = f(\omega)
+ {\bar f} (\bar \omega) \;\;\;,
\label{linm}
\eeqa
where $\omega = a \ln z = a ( \ln r + i \theta ) = {\hat r} + i 
{\hat \theta}$.
Here, $a = \frac{n+1}{2 \pi} \sqrt{|\alpha_i \alpha_{i + 7}|}$, where
$\alpha_i$ and $\alpha_{i + 7}$ denote two electric charges
carried by each of the solutions in this class.  The associated
field strengths of the three-dimensional gauge fields are 
\cite{BC}
\beqa\label{fs1}
F_{t \beta}^{(1)i} = \eta_{\alpha_i} \sqrt{G^{ii}} \;
\frac{\partial_{\beta}H}{H^2}\;\;\;,\;\;\;
F_{t \beta \,i}^{(2)} = - \eta_{\alpha_i} \sqrt{G_{ii}} \;
\frac{\partial_{\beta}H}{H^2} \;\;\;,\;\;\; \beta = {\hat r}, {\hat
  \theta} \;\;\;,\;\;\;  \eta_{\alpha_i} = \pm \;.
\label{gf2}
\eeqa
The three-dimensional dilaton field is given by $e^{- \phi} = H$.
The i-th component of the internal metric $G_{mn}$ reads
$G_{ii} = |\frac{\alpha_i}{\alpha_{i+7}}|$.  In addition, there are
(depending on the integer $n$) various additional non-constant background 
fields $G_{mn}$ and $B_{mn}$, which are also determined in terms of
$f(\omega)$ \cite{BC}.

Solving the Killing spinor equations (\ref{killing}) does not determine
the form of $f(\omega)$.  Its form can be determined by
demanding the solution to behave as 
$f(\omega) \approx \frac{\omega}{2(n+1)}$
at spatial infinity \cite{BBC}
and by requiring the solution to have finite energy, as follows \cite{BC}.  
The energy carried by any of the solutions in this class is compute
to be (in units where $8 \pi G_N =1$) 
\beqa
E &=&i \, 2n 
\int d\omega d{\bar \omega}\, \frac{\partial_{\omega}f \partial_{\bar \omega}
{\bar f}}
{(f + {\bar f})^2}
= i \, 2n \int dz d{\bar z}\, 
\frac{\partial_{z} {\hat f} 
\partial_{\bar z}{\bar {\hat f}}}
{({\hat f} + {\bar {\hat f}})^2} \;\;\;,
\label{enint}
\eeqa
where we have introduced ${\hat f} =
\frac{n+1}{\pi}f$ for later convenience.  
There is an elegant mechanism \cite{GSVY}
for rendering the integral (\ref{enint})
finite.  Let us take $z$ to be the coordinate of a complex plane.  Then
there is a one-to-one map from a certain domain F on the ${\hat f}$-plane
(the so called `fundamental' domain) to the $z$-plane.  This map is known
as the j-function, $j({\hat f}) = z$.  By means of this map, the integral
(\ref{enint}) can be pulled back from the $z$-plane to the domain F
(the $z$-plane covers F exactly once).  Then, by using integration by
parts,
this integral can be related to a line integral over the boundary of F,
which is evaluated to be \cite{GSVY} 
\beqa
E = 2n\,\frac{2 \pi}{12} = 2n \,\frac{\pi}{6}
\label{ener2}
\eeqa
and, hence, is finite.
By expanding $j({\hat f}) = 
e^{2 \pi {\hat f}} + 744 + {\cal O} ( e^{-2 \pi {\hat f}} ) = z$ we 
recover
$f(\omega) \approx \frac{\omega}{2(n+1)}$ at spatial infinity.

We note that the solutions discussed above represent one-center
solutions.  They can be generalised to multi-center solutions via
$j({\hat f}(z)) = P(z)/Q(z)$, where $P(z)$ and $Q(z)$ are polynomials in
$z$ with no common factors.  These are the analogue of the multi-string
configurations discussed in \cite{GSVY}.

It can be checked that the curvature scalar ${\cal R} \propto
\partial_{\omega} f \partial_{\bar \omega} {\bar f}$ 
blows up at the special point
${\hat f}=1$ (at this point, the $j$-function and its derivatives are given
by $j = 1728, j' =0$), 
whereas it is well behaved at the point 
${\hat f} = 
e^{i \pi/6}$  (at this point, the $j$-function and its derivatives are given
by $j  = j'= j''=0$).  It would be interesting to understand the
physics at this special point in moduli space further.

As mentioned in the introduction the three-dimensional solutions
reviewed in this section can be given an eleven dimensional
interpretation \cite{BC} in terms of orthogonally intersecting
M-branes.  The two gauge fields (\ref{gf2}), in particular, arise 
from a wave and from a M-2-brane in eleven dimensions, respectively.

\section{Wrapped M-monopoles}

Another class of solutions constructed in 
\cite{BC} is the class of solutions carrying energy $E = n \frac{\pi}{6}$,
where $n=1,2,3,4$.  An example of a solution
carrying energy $E = \frac{\pi}{6}$ is obtained by 
compactifying the M-monopole in the following way.  The M-monopole 
in eleven dimensions is given by the metric 
\cite{sorkin}
\eq\label{sork}
ds_{11}^2=-dt^2 + Hdy_i^2 + H^{-1}( d\psi+A_idy_i)^2 + dx_1^2 + \dots +
dx_6^2,\qquad  i=1,2,3,
\en
with $H=H(y_i),\,\,\, F_{ij}=\del_iA_j - \del_j A_i =
c\,\varepsilon_{ijk}\del_k H, \,\,\, c=\pm.$  Here, $\psi$ denotes a periodic
variable.  Identifying one of the $x_i$ with the eleven-th cooordinate
and compactifying the remaining $x_i$ on a five-torus yields the 
five-dimensional line element
\eq
ds_5^2 = -dt^2 + Hdy_i^2 + H^{-1}( d\psi + A_idy_i)^2.
\label{sork5}
\en
Now, consider the case that $H$ only depends on two of the coordinates
$y_i$, that is $H=H(z, \bar z)$ with $z = y_2 + i y_3$.  Then we can set
$A_2=A_3=0$, and $\del_2 A_1 =
-c\del_3 H,\,\del_3 A_1=c\del_2 H.$ 
The metric is now
\eqn
ds_5^2 &=& -dt^2 + H dz d{\bar z} + H dy_1^2 + H^{-1}( d \psi
+ A_1dy_1)^2 \no\\
       &=& g_{\mu\nu} dx^\mu dx^\nu + G_{mn}dx^mdx^n,
\enq
where
the off-diagonal internal metric is given by
\eq\label{km}
\ G_{mn}=  \left( \begin{array}{cl}
H+A_1^2 H^{-1} &  A_1H^{-1} \\
 A_1H^{-1} & H^{-1}  
\end{array} \right)\;\;.
\en
The resulting three-dimensional solution can be turned into a finite
energy solution with energy $E = \frac{\pi}{6}$ by using the mechanism
 described in the previous
section.

Let us now compare this solution to the one obtained by identifying
the eleven-th dimension not with one of the $x_i$, but rather
with one of the $y_i$ \cite{BlauLoHull}.  Compactifying the eleven-dimensional
line element (\ref{sork}) on a six-torus yields again the five-dimensional
line element (\ref{sork5}).  Compactifying over $\psi$ yields the
four-dimensional line element
\eq
ds_4^2 = -dt^2 + Hdy_i^2 
\en
with internal metric component $G_{\psi \psi} = H^{-1}$ 
and magnetic gauge field $F_{ij}=
c\,\varepsilon_{ijk}\del_k H$.  The Bianchi identity of $F$ yields
$\Delta H =0$.  Now, identifying one of the transverse coordinates
$y_i$ with the eleven-th 
coordinate (say $y_1 = x_{11}$) and setting $z= y_2 + i y_3$ yields
the Laplacian as $\Delta = \partial^2_{x_{11}} + 
4 \partial_z \partial_{\bar z}$.  Thus, $H$ is now a harmonic function on 
$S^1/Z_2 \times {\rm R^2}$.  In \cite{vafoog} a 
solution to the Laplace equation on $S^1 \times {\rm R^2}$ 
was constructed.  It can be readily adapted to the case at hand.
We take the range of $x_{11}$ to be $x_{11} \in [ - \frac{1}{2}, \frac{1}{2}]$,
with a peridic identification $x_{11} \sim x_{11} + 1$ of the endpoints.
The $Z_2$ symmetry acts as $x_{11} \rightarrow - x_{11}$.
A $Z_2$ invariant solution 
to the Laplace equation on $S^1/Z_2 \times {\rm R^2}$
is then given by 
\beqa
H (x_{11}, z, {\bar z}) = \sum_{n \in {\rm \bf Z}} \left(
\frac{1}{|n - 
\frac{1}{2}|} - \frac{1}{\sqrt{(x_{11} - 
(n - \frac{1}{2}) )^2 + z {\bar z}}} \right) \;\;\;.
\label{hvo}
\eeqa
The constant
term in (\ref{hvo}) is such that $H$ is regular at $z=0$ on the orbifold
plane $x_{11} =0$.  For $|z| \rightarrow \infty$ $H$, on the other hand,  
$H$ reduces to $H \approx \log z \bar z$ on the plane $x_{11}=0$.

\bigskip
   
\section*{ Acknowledgement}  

We would like to thank K. Behrndt, M. Bourdeau and E. Verlinde for 
useful discussions.



\begin{thebibliography}{99}










\bibitem{DuRaLu} M. J. Duff, R. R. Khuri and J. X. Lu, {\it Phys. Rep.} 
{\bf 259} (1995) 213, hep-th/9412184



\bibitem{Gaunt} J. P. Gauntlett, hep-th/9705011

\bibitem{ts} A. A. Tseytlin, {\it Nucl. Phys.} 
         {\bf B475} (1996) 149, hep-th/9604035


\bibitem{berg1}  E. Bergshoeff, M. de Roo, E. Eyras, B. Janssen and J. P. van
         der Schaar, {\it Class. Quant. Grav.} {\bf 14} (1997) 2757, 
       hep-th/9704120


\bibitem{sen1} A. Sen, {\it Nucl. Phys.} {\bf B434} (1995) 179, 
          hep-th/9408083

\bibitem{BBC} I. Bakas, M. Bourdeau and G. L. Cardoso, hep-th/9706032

\bibitem{BC} M. Bourdeau and G. L. Cardoso, hep-th/9709174

\bibitem{BlauLoHull} M. Blau and M. O'Loughlin, hep-th/9712047 \\
          C. M. Hull, hep-th/9712075


\bibitem{ElGiKuRa}  S. Elitzur, A. Giveon, D. Kutasov and E. Rabinovici, 
              {\it Nucl. Phys.} {\bf B509} (1998) 122,      
              hep-th/9707217

\bibitem{GSVY} B. R. Greene, A. Shapere, C. Vafa and S.--T. Yau, 
      {\it Nucl. Phys.} {\bf B337} (1990) 1

\bibitem{mah} S. Hassan and A. Sen, {\it Nucl. Phys.} {\bf B375} (1992) 103, 
               hep-th/9109038 \\
         J. Maharana and J. H. Schwarz, {\it Nucl. Phys.} 
            {\bf B390} (1993) 3,
              hep-th/9207016 






\bibitem{sorkin} R. D. Sorkin, {\it Phys. Rev. Lett.} {\bf 51} (1983) 87 \\
        D. J. Gross and M. J. Perry, {\it Nucl. Phys.} {\bf B226} (1983) 29 



\bibitem{vafoog} H. Ooguri and C. Vafa, {\it Phys. Rev. Lett.} {\bf 77} (1996)
                    3296, hep-th/9608079

\end{thebibliography}
\end{document}